\documentclass[conference]{IEEEtran}

\newcommand{\emaildot}{\makebox[0.2em]{\scalebox{.23}{\textbullet}}}
\usepackage{hyperref}
\hypersetup{draft}

\usepackage[font=small,skip=10pt]{caption}
\usepackage{cite}
\usepackage{caption,subcaption}
\usepackage{dblfloatfix}
\usepackage{algorithm}
\usepackage{algorithmic}
\usepackage{RobStd}
\usepackage{multirow}
\usepackage{paralist}
\usepackage{cprotect}
\usepackage{amssymb}
\usepackage{dirtytalk}
\usepackage{amsmath}
\usepackage{booktabs} 
\usepackage{subcaption}
\usepackage{graphicx}
\usepackage{svg}
\usepackage{xcolor}
\usepackage{siunitx}
\usepackage{todonotes}
\usepackage{pgfplots}
\usepackage{pgfpie3}
\usepackage{bchart}
\usepackage[eulergreek]{sansmath}
\usepackage{helvet}
\usetikzlibrary{arrows}

\definecolor{mpink}{HTML}{ef476f}
\definecolor{myellow}{HTML}{ffd166}
\definecolor{mgreen}{HTML}{06d6a0}
\definecolor{mblue}{HTML}{118ab2}
\definecolor{mdarkblue}{HTML}{073b4c}

\definecolor{ared}{HTML}{ff898d}
\definecolor{ayellow}{HTML}{ffda74}
\definecolor{ablue}{HTML}{afe160}
\definecolor{agreen}{HTML}{4bace8}
\definecolor{apurple}{HTML}{967bbb}

\definecolor{modern_green}{HTML}{70AD47}
\definecolor{modern_blue}{HTML}{00B0F0}
\definecolor{modern_lightblue}{HTML}{A1DFF6}
\definecolor{modern_yellow}{HTML}{FFC000}
\definecolor{modern_red}{HTML}{E42831}
\definecolor{modern_darkgray}{HTML}{434343}
\definecolor{modern_darkblue}{HTML}{0083b2}
\definecolor{modern_lred}{HTML}{F2999D}
\definecolor{modern_lgreen}{HTML}{c5e384}
\definecolor{modern_lyellow}{HTML}{fce883}
\definecolor{modern_violet}{HTML}{967bbb}

\definecolor{moderngreen}{HTML}{70AD47}
\definecolor{modernblue}{HTML}{00B0F0}
\definecolor{modernlightblue}{HTML}{A1DFF6}
\definecolor{modernyellow}{HTML}{FFC000}
\definecolor{modernred}{HTML}{E42831}
\definecolor{moderndarkgray}{HTML}{434343}
\definecolor{moderndarkblue}{HTML}{0083b2}
\definecolor{modernlred}{HTML}{F2999D}
\definecolor{modernlgreen}{HTML}{c5e384}
\definecolor{modernlyellow}{HTML}{fce883}
\definecolor{modernviolet}{HTML}{967bbb}

\definecolor{blue1}{HTML}{03045e}
\definecolor{blue2}{HTML}{023e8a}
\definecolor{blue3}{HTML}{0077b6}
\definecolor{blue4}{HTML}{0096c7}
\definecolor{blue5}{HTML}{00b4d8}
\definecolor{blue6}{HTML}{48cae4}
\definecolor{blue7}{HTML}{90e0ef}
\definecolor{blue8}{HTML}{ade8f4}
\definecolor{blue9}{HTML}{caf0f8}

\renewcommand{\sfdefault}{phv} 
\pgfplotsset{
  compat =1.17,
  tick label style = {font=\sansmath\sffamily\scriptsize},
  every axis label = {font=\sansmath\sffamily\scriptsize},
  legend style = {font=\sansmath\sffamily\scriptsize},
  label style = {font=\sansmath\sffamily\scriptsize},
}

\newcommand\notes[1]{\textcolor{black}{#1}}
\newcommand\notesdone[1]{\textcolor{black}{}}

\newcommand\dayane[1]{\textcolor{black}{#1}}
\newcommand\chesca[1]{\textcolor{black}{#1}}

\graphicspath{{images/}}

\newcommand{\rs}{RecSys\xspace}
\newcommand{\fw}{iMARS\xspace}
\newcommand{\et}{ET\xspace}
\newcommand{\ets}{ETs\xspace}

\begin{document}

\title{iMARS: An In-Memory-Computing Architecture for Recommendation Systems}


\author{
\IEEEauthorblockN{Mengyuan Li, Ann Franchesca Laguna, Dayane Reis, Xunzhao Yin\IEEEauthorrefmark{1}, Michael Niemier, and Xiaobo Sharon Hu}

\IEEEauthorblockA{\IEEEauthorrefmark{1}College of Information Science and Electronic Engineering, Zhejiang University, China 310027, xzyin1@zju.edu.cn}
\IEEEauthorblockA{Department of Computer Science and Engineering\\
University of Notre Dame, Notre Dame, IN, USA, 46556\\
\{mli22, alaguna, dreis, mniemier, shu\}@nd\emaildot edu}}

\maketitle
\begin{abstract}

Recommendation systems (\rs) suggest items to users by predicting their preferences based on historical data. Typical \rs handle large embedding tables and many embedding table related operations. The memory size and bandwidth of the conventional computer architecture restrict the performance of \rs. This work proposes an in-memory-computing (IMC) architecture (\fw) for accelerating the filtering and ranking stages of deep neural network-based \rs. \fw leverages IMC-friendly embedding tables implemented inside a ferroelectric FET based IMC fabric. Circuit-level and system-level evaluation show that \fw achieves 16.8$\times$ (713$\times$) end-to-end latency (energy) improvement compared to the GPU counterpart for the MovieLens dataset.


%
\end{abstract}


%

\section{Introduction}
\label{sec:introduction}



Recommendation systems (\rs) have been used in various applications to suggest items such as movies, music, books, shopping items, websites, etc. based on a user's previous behavior, as well as historical data from other users. The large amount of available data allows \rs to leverage deep neural networks (DNN). DNN-based \rs have been widely adopted by companies like Facebook~\cite{DLRM19} and Google~\cite{covington2016deep} to improve their online services. 





 The typical DNN-based \rs inference process~\cite{covington2016deep} includes two stages: \textit{filtering} and \textit{ranking}. In the filtering stage, the system selects a set of candidate items from a large item database to be recommended based on a user's behavior. The ranking stage then computes the probability of choosing each candidate item. The items with the highest probability are finally returned to the user. Both the filtering and ranking stage  use embedding tables (ETs) to capture and store user behaviors and item characteristics. The large \ets make the operations on them memory-bandwidth limited. 

Several algorithm and hardware solutions, e.g., \et compression \cite{shi2020compositional}, have been proposed to alleviate the memory-bandwidth  bottleneck. Near-memory computing has also been leveraged to solve the memory-bandwidth problem by bringing the compute units closer to the memory \cite{ke2020recnmp}. However, these solutions only focus on the ranking stage and not the filtering stage which are still impeded by memory bottlenecks due to the huge amount of data transfers.

In-memory-computing (IMC) is a computational paradigm that can alleviate the data transfer overhead between the memory and the compute unit by performing logic and arithmetic operations inside the memory unit itself. Different IMC kernels have been proposed, such as content addressable memories (CAMs), crossbars, general purpose computing-in-memory (GPCiM) and configurable memory arrays (CMAs). CAMs \cite{ni2019ferroelectric} can perform parallel content-based searches in the memory itself, crossbar arrays \cite{ranjan2019x} can perform matrix-vector multiplications, and GPCiM \cite{reis2018computing} performs Boolean logic  and  arithmetic  operations in memory. CMAs combine the functionalities of random access memory (RAM), CAM, GPCiM and crossbar in a single memory array. Ferroelectric field-effect transistors (FeFET) based crossbars, CAMS, GPCiM and CMAs have been designed \cite{zhang19_femat, reis2020_aim} and have shown that FeFET-based circuits are  denser and faster than CMOS-based and ReRAM-based circuits. FeFETs can be easily integrated with the CMOS fabrication process and large-scale FeFET memories have also been fabricated in ~\cite{ dunkel17}. 

In this paper, we propose \fw, an IMC architecture for \rs. \fw exploits an FeFET-based CMA fabric that combines the functionalities of RAM, ternary CAMs (TCAMs) and GPCiM. The FeFET-based CMA (from \cite{reis2020_aim}) can switch its functionality to implement (i) TCAM-based searches to realize nearest neighbor search (NNS) in the filtering stage; and (ii) GPCiM-based arithmetic logic to implement the additions/accumulations in the filtering and ranking stage. \textbf{Specific contributions of our work include (1) an integrated IMC fabric to simultaneously accelerate the filtering and ranking stages; (2) an IMC-friendly computation flow to facilitate mapping the RecSys algorithms to \fw; (3) support for all \et related operations {\em in memory} by combining TCAM and GPCiM functionality in the CMA fabric; (4) a two-level memory hierarchy and corresponding in-memory adder trees to store the large ETs.} 

\notes{We have evaluated the latency and energy benefits of \fw based on two widely used RecSys models: YoutubeDNN~\cite{covington2016deep} on the MovieLens dataset~\cite{harper2015movielens} and Facebook DLRM~\cite{DLRM19} on the Criteo Kaggle dataset. The results show that for the MovieLens dataset, \fw  achieves a $16.8\times$ (713$\times$) end-to-end speedup (energy improvement) against the GPU implementation. For the Criteo Kaggle dataset which is widely used for the ranking task, the ranking model in Facebook DLRM is accelerated with \fw, which leads to 13.2$\times$ (57.8$\times$) improvement in latency (energy) improvement.}



\section{Background}
\label{sec:background}


In this section, we review the basics of DNN-based \rs and related work on hardware accelerators for \rs. Furthermore, we discuss the IMC circuits (TCAMs, GPCiMs, CMAs and crossbars) and technologies (i.e., CMOS, FeFETs) employed in \fw.

\subsection{Recommendation Systems}
\label{sec:rsbasics}

\begin{figure}
    \centering
    \includegraphics[width=0.4\textwidth]{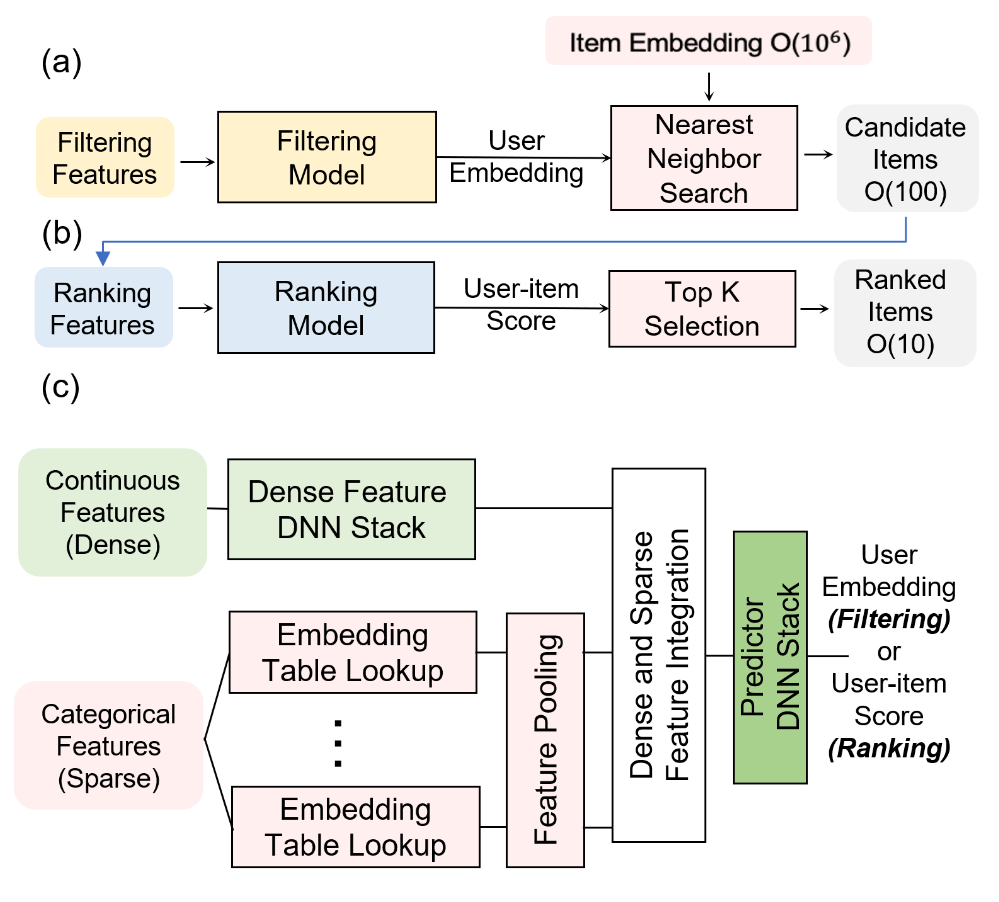}
    \vspace{-2ex}
    \caption{Configuration of the (a) filtering and (b) ranking stages. (c) General DNN model used in the filtering and ranking stages.}
    \vspace{-2ex}
    \label{fig:Algorithm}
\end{figure}

\rs are composed of a filtering and a ranking stage \cite{covington2016deep}. The filtering stage (Fig.~\ref{fig:Algorithm}(a)) aims to reduce the number of computations needed in the ranking stage by determining a set of candidate items (e.g. $O(100)$) from the entire item database (e.g. $O(10^6)$). The ranking stage (Fig.~\ref{fig:Algorithm}(b)) aims to find the item with the highest score for a specific user from the candidate items. The filtering stage uses a DNN to characterize the user behavior as a single embedding vector. Based on this user behavior, the candidate items to recommend can be found by using the NNS on the item ET as shown in Fig.~\ref{fig:Algorithm}(a). The goal of the ranking stage is to evaluate each user-item pair and predict the score of each candidate item for a specific user. The score is defined as the click-through rate (CTR) and indicates the likelihood that an item will be clicked. Based on the scores, the top-$k$ items ($O(10)$) are returned to the user (Fig.~\ref{fig:Algorithm}(b)). 

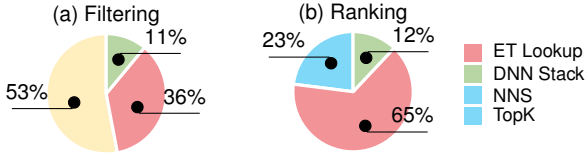
\begin{figure}
    \sansmath\sffamily 
    \raggedright
    \begin{subfigure}[b]{0.15\textwidth}
    \begin{tikzpicture}
        [
        pie chart,
        slice type={pieblue}{modern_yellow!25},
        slice type={piered}{modern_red!50},
        slice type={piegreen}{modern_green!50},
        pie values/.style={font={\footnotesize}},
        pie title/.style={font={\footnotesize}},
        scale=0.8,
    ]
      \pie{(a) Filtering}{53/pieblue,36/piered,11/piegreen}
    \end{tikzpicture}
    \end{subfigure} \hspace{0.5cm}
    \begin{subfigure}[b]{0.15\textwidth}
    \begin{tikzpicture}
        [
        pie chart,
        slice type={piered}{modern_red!50},
        slice type={pieblue}{modern_blue!50},
        slice type={piegreen}{modern_green!50},
        slice type={pieyellow}{modern_blue!50},
        pie values/.style={font={\footnotesize}},
        pie title/.style={font={\footnotesize}},
        scale=0.8,
    ]
      \pie{(b) Ranking}{23/pieyellow,65/piered,12/piegreen}
      \legend[shift={(2cm,1cm)},draw opacity=0]{{ET Lookup}/piered,{DNN Stack }/piegreen, {NNS}/pieblue, {TopK}/pieyellow}
    \end{tikzpicture}
    \end{subfigure}

    \caption{Operation breakdown of the filtering and ranking stages on the MovieLens dataset.
    }
    \vspace{-1ex}
    \label{fig:breakdown2}
\end{figure}

Both DNN models in the filtering and ranking stages follow the configuration depicted in Fig.~\ref{fig:Algorithm}(c).
DNNs are employed to generate the embedding vector representing the user behavior or predicting user-item pair scores. The models take advantage of both continuous (dense) and categorical (sparse) features. Dense features can be directly processed by a DNN while sparse features are captured by large \ets with sparse lookup and pooling operations. 
These \et operations (lookups, NNS) contribute a significant portion of the run time in \rs as shown in the operation breakdown of the MovieLens dataset using the YouTubeDNN \rs \cite{covington2016deep} in Fig. \ref{fig:breakdown2}.

\label{sec:hardware_for_rs_related_work}



Existing efforts on accelerating \rs include field programmable gate array (FPGA)-based accelerators. E.g., FleetRec~\cite{FleetRec} and MicroRec~\cite{MicroRec} alleviate the memory-bandwidth bottleneck of the \ets in \rs by using FPGAs with high bandwidth memory.
Near-memory computing has also been considered to alleviate the memory bottleneck. RecNMP~\cite{ke2020recnmp} uses a dual in-line memory module-based near-memory computing that can support sparse embedding models.
Most of these existing hardware accelerators only focus on a single aspect of \rs, i.e., ranking, filtering, DNNs or \ets. Though it is possible to simply cascade these accelerators to implement \rs, such a simple-minded approach results in higher hardware cost due to duplicated components.



\subsection{In-Memory Computing Circuits}\label{sec:IMC}

\begin{figure*}
    \centering
    \includegraphics[width=0.9\linewidth]{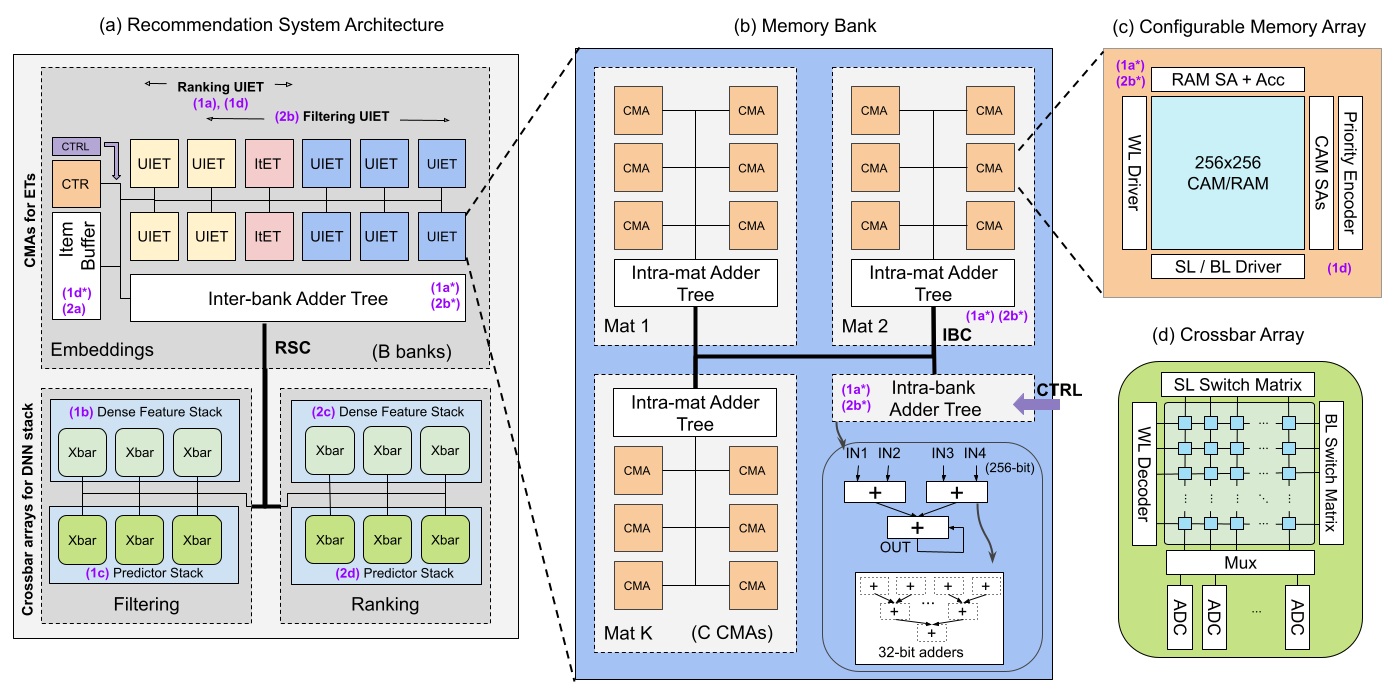}
    \caption{(a) Proposed \fw architecture. (b) CMA bank for item embedding table (ItET) and user-item embedding table (UIETs). (c) CMA structure, which is capable of switching between the CAM/RAM/GPCiM modes. (d) Crossbar array.  Labels (1a), ..., (2d*) shows the computation flow of the ranking and filtering stages as discussed in Sec. \ref{method:mapping}.}
    \label{fig:archmapping}
    
\end{figure*}

IMC circuits can alleviate the memory bottleneck and provide parallelism to \rs operations. This section discusses different types of IMC circuits such as TCAMs, crossbars, GPCiMs and CMA.

\textbf{TCAMs} enable parallel searches based on the Hamming distance for a query against a large stored database in $O(1)$ time \cite{ni2019ferroelectric}. Using the threshold-match mode of the TCAMs, we can retrieve the row entries in the array nearest to the query under the threshold distance by parallel searches. In a TCAM array with $r$ rows and $c$ columns, the TCAM cells in a row are serially connected through a common matchline. Each TCAM cell, $s_{ij}$, performs an XOR operation between bit $j$ of query $Q$ and the bit $j$ stored at $s_{ij}$. Each matchline implements a logic AND 
of all the cells connected to it.


\textbf{Crossbars} are an IMC structure where every input is connected to every output through cross-points that consist of memory elements and selectors. Crossbars can efficiently implement matrix-vector multiplications, and are thus ideal accelerators for DNN models such as convolutional neural networks \cite{ISAAC}. 


\textbf{GPCiMs} \cite{jeloka16, reis2018computing} perform general-purpose Boolean logic and arithmetic operations inside RAM. GPCiMs can employ specialized memory cells based on emerging technologies to perform current-based operations inside a memory array (e.g., \cite{reis2018computing}). Alternatively, customized memory peripherals (such as sense amplifiers) can be designed to operate with two memory words that are simultaneously selected by row decoders. In GPCiMs that employ customized sensed amplifiers, the voltage (or current flow) through a column-connected bitline is sensed and compared to one (or multiple) reference(s) so the results of Boolean logic/arithmetic can be produced.



\textbf{CMAs} \cite{jeloka16, zhang19_femat, reis2020_aim} combine multiple IMC functionalities in the same physical structure. For instance, CMAs can work as either TCAM or GPCiM units at distinct times. Note that conventional TCAMs perform row-wise sensing as matchlines are placed along the row direction. GPCiMs, on the other hand, require the voltage drop (or current flow) through the vertically-connected bitlines to be sensed and compared to one (or more) reference(s) in order to produce the results needed for general-purpose computation. Due to this difference in the  TCAM and GPCiM arrays, combining them in a single hardware structure requires additional memory peripherals to achieve re-configurability \cite{zhang19_femat, reis2020_aim}. 


Storing \rs \ets requires significant amount of memory and substantial communication between memory and processing units, which could be alleviated by IMC architectures. The use of CMAs based on emerging technologies can be beneficial for implementing IMC-based \rs compared to standard CMOS CMAs \cite{jeloka16} due to the increased density of memory cells and lower standby power (a result of the device's non-volatility). 

FeFETs have been used in various IMC based circuits such as CAMs, GPCiMs and CMAs \cite{chen2019impact, beyer20}.
Previous work has demonstrated the benefits of FeFET-based CMAs over other emerging technologies such as ReRAMs \cite{zhang19_femat, reis2020_aim}. 
FeFETs have similar structure as metal-oxide-semiconductor field-effect transistors (MOSFETs) used in standard CMOS silicon, except a layer of FE oxide is deposited in the transistor's gate stack. Because of this,  FeFETs are compatible with the CMOS fabrication process ~\cite{ dunkel17} and large-scale FeFET memories have been demonstrated~\cite{ dunkel17}. For these reasons, we employ a FeFET-based CMA design for accelerating \rs. 





\section{IMC Accelerator for Recommendation Systems} \label{sec:imrs}

\chesca{We propose \fw, an architecture for \rs that uses an IMC fabric to accelerate the DNN stack and \et related operations in both the filtering and ranking stage of \rs. The \rs ranking and filtering stages use sparse and dense features in determining which items to recommend to the user. The dense features are sent to the DNN stack while the sparse features are sent to \ets.} For the DNN stack, crossbars can be readily leveraged. For \et related operations, \dayane{we leverage CMAs and adder trees placed in the memory periphery.} Since different \ets play different roles depending on the stage, how to organize these tables \dayane{inside the IMC fabric} must be considered carefully.  We elaborate our proposed architecture in Sec. \ref{method:arch} and computation mapping in Sec. \ref{method:mapping}.







\subsection{\fw Architecture} \label{method:arch}

The \fw architecture, shown in Fig. \ref{fig:archmapping}(a), consists of two types of IMC arrays: Crossbar arrays for the DNN stack (bottom) and CMA arrays for the \ets (top). The implementation of the DNN and the \ets inside \fw is described below. 

\subsubsection{Embedding Tables}

\chesca{Some sparse features are used in both the ranking and filtering stage and hence can share the same \ets. We employ two different \ets: user-item embedding table (UIET) and the item embedding table (ItET). The UIET holds user-item features used by the filtering and ranking stage. Some UIETs are exclusively used by the filtering or the ranking stage while some UIETs are shared by the two stages. The item characteristics are stored in the ItET. To conserve memory space, the ItET can be accessed by both the filtering and the ranking stage.}

For the UIETs, \et lookups and pooling operations are done in the filtering and ranking stage, respectively. For the ItET, two primary operations are required: (i) lookups and pooling on the \et to transform a sparse input feature into a dense vector; (ii) the NNS on the \et to return the candidate item IDs in the filtering stage. The \fw architecture implements the two types of \ets inside its IMC fabric. CMAs integrate reads, searches, and in-memory logic/arithmetic operations in a single array, which makes them a suitable choice for implementing UIETs and ItETs.

\dayane{While the circuits inside a CMA are described in \cite{reis2020_aim}, we introduce the design of a novel, two-level hierarchy based on CMAs and adder trees to store and efficiently compute on the large number of items in the \ets as needed by the \rs. The components that implement \ets inside \fw (i.e., the CMAs and adder trees) are discussed below.}


\dayane{\textbf{Hierarchical CMAs:} To accomodate the large \ets, $B$ banks of CMAs are deployed in \fw. Fig. \ref{fig:archmapping}(b) depicts the structure of one CMA bank, which consists of $M$ mats (labeled as Mat-1, Mat-2, ..., Mat-$M$). Each mat is comprised of $C$ CMAs that work independently as the IMC engines in \fw for performing lookups, searches and additions. An individual CMA is depicted in Fig. \ref{fig:archmapping}(c). It employs CAM sense amplifiers (SAs) based on a preset threshold, as well as searchline (SL) drivers and priority encoders to perform threshold based NNS. We chose to implement threshold based matching in the CAM based on a reference current generated by a dummy 1T+1FeFET cell, which can be adjusted to compensate for process variations or to change the sensitivity of the Hamming distance in the NNS operation. The RAM SAs, wordline (WL) and bitline (BL) drivers are used during lookups. Pooling operations are performed with in-memory additions (through an accumulator placed next to the RAM SA). More details on the CMA structure can be found in \cite{reis2020_aim}. } \dayane{FeFET-based CMAs \cite{reis2020_aim} are utilized to implement the \ets. When compared to CMOS-based counterparts, FeFET-based CMAs have higher density (which helps to reduce the area footprint) and lower leakage power \cite{zhang19_femat}.}


\dayane{\textbf{Adder trees:} To support the accumulation of a large number of parameters, we develop a hierarchical, adder tree structure. Specifically, \fw uses in-memory addition to sum, in a single memory array, embeddings comprised of 32 dimensions with int-8 quantization. To accumulate (sum up) the outputs of the CMAs for each mat, \fw sums up $C$ 256-bit (i.e., 32$\times$8)-bit numbers  leveraging a near-memory, 256-bit intra-mat adder tree placed in each mat. Different mats can perform intra-mat additions in parallel. Once the intra-mat additions are completed, their results are accumulated across the $K$ mats to produce a single 256-bit result (one output per memory bank). \fw supports the addition of four 256-bit inputs inside a near-memory ``Intra-bank Adder Tree" in one shot (bottom right of Fig. \ref{fig:archmapping}(b)). In other words, we design an ``Intra-bank Adder Tree" with a fan-in of 4, a design choice made as a compromise between area footprint of the \fw banks and performance of the intra-bank addition. In cases where more inputs need to be accumulated (i.e., when $K>4$), multiple rounds of addition are needed using the same ``Intra-bank Adder Tree". }

\dayane{The aforementioned design parameters $B$, $M$ and $C$ largely impact the area, capacity and the performance of \fw. First, area footprint increases proportionally to $B$, $M$ and $C$. Larger $B$, $M$ and $C$ increase the capacity of \fw for storing embeddings. High capacity enables \fw to accomodate big workloads. However, a large $C$ implies a large fan-in for the ``Intra-mat Adder Tree" inside each mat, which leads to parasitic effects that increases the delay for aggregating the outputs of multiple CMAs. A large $K$, in turn, implies that more mats are connected to (and sharing) the same communication bus, which increases the overall latency of the \rs (to be discussed in Sec. \ref{method:comms}). }

\subsubsection{DNN Stack}

\chesca{The DNN stack requires matrix-vector multiplications which can be implemented using crossbar arrays in \fw (Fig. \ref{fig:archmapping}(a))}. Two dedicated crossbar banks are employed to execute the ranking and the filtering DNN stack composed of fully connected layers. Each crossbar bank contains multiple crossbar arrays (Fig. \ref{fig:archmapping}(d)) in order to accomodate the respective DNN model. \chesca{These crossbar banks both hold the DNN Stack to obtain dense features and the DNN stack that returns a user embedding during the filtering stage or a user item-score for the ranking stage.} Crossbar arrays can leverage the FeFET technology \cite{Neurosim}.

\subsubsection{Communication inside \fw} \label{method:comms}
Data among the different hardware components of \fw need to be communicated. To ensure such communication does not incur too much overhead, we carefully design communication channels inside \fw. There are two types of communication in \fw: (1) communication among the  different functional blocks and (2) communication between the mats in each CMA bank. While (1) leverages the \rs communication (RSC) bus, (2) occurs through the intra-bank communication (IBC) network. The RSC bus and the IBC network are depicted in Fig. \ref{fig:archmapping}(a) and (b), respectively.

The RSC bus enables the exchange of inputs/outputs through the different hardware blocks in \fw. While data traffic between the different hardware blocks is essential to the system's overall functionality, the data traffic through the RSC bus is not as intense as the traffic inside the memory banks that store the \ets. Data communication on the IBC network and the RSC bus is serialized to minimize the wiring overhead (thus, reducing the area of \fw). \fw is designed for a RSC bus with 256-bit capacity. The IBC, on the other hand, supports the transmission of  128 bytes of data (i.e., four 256-bit inputs) to be added up inside the intra-bank adder tree (bottom right of Fig. \ref{fig:archmapping}(b)) in one shot. \dayane{Data communication on the IBC network is serialized when $K>$4 (the fan-in of the ``Intra-bank Adder Tree"). The IBC network capacity and the number of ``Intra-bank Adder Tree" fan-ins are design choices that must take into consideration the impact on area footprint, delay and energy. For instance, extremely wide buses may be impractical as they require too much area to be implemented. On the other hand, a narrow IBC would require many data fetches through the bus across $K$ mats. The overhead of communication with the RSC bus/IBC network is accounted for in the results reported in Sec. \ref{sec:evaluation}.}

\dayane{Data traffic inside the RSC and the IBC is orchestrated by a controller circuit (indicated by the box labeled CTRL in Fig. \ref{fig:archmapping}(a)). The controller circuit consists of a clock generator and two counters that keep track of (i) the activated bank, and (ii) the mats inside the bank that are sending outputs for accumulation with the ``Intra-bank Adder Tree" circuit. Data packets always travel through the IBC in a predetermined order, as defined by the counters (i.e., in Bank $B$, from Mat-1, Mat-2, ..., Mat-$M$ in groups of four outputs, i.e., 128 bytes). The pre-defined communication pattern reduces conflicting accesses and eliminates the need for routers.}



\subsection{Embedding Table Mapping}

\chesca{The \fw architecture uses CMAs to store \ets., which have varying number of entries (typically 3-30,000 entries). Each row on the CMA represents an entry of an \et. Designing CMAs with varying sizes is not practical. We determined the optimal array-level CMA to be the size of $256\times256$ based on circuit-level simulations. Some of the \ets can fit in a single CMA and some require multiple CMAs. The number of CMAs needed to store an \et is $n/R$ where $n$ is the number of entries in the \et and $R$ is the number of rows in the CMA. If $n/R<C$, we only need one mat, otherwise the number of mats needed to be activated is equal to $n/(RC)$. 
Each sparse feature is mapped to a separate bank. Hence, the number of activated banks depends on the number of sparse features. }

We quantize all \ets to 8-bit integer precision to reduce the memory requirement. We also replace the cosine-distance based NNS in the original filtering stage with the IMC-friendly Hamming-distance based NNS. To facilitate the Hamming distance search, we employ a locality-sensitive hashing (LSH) technique on the ItET~\cite{ni2019ferroelectric}. Each row of the ItET includes the additional bits for storing the corresponding LSH values. Finally, a \textit{fixed-radius near neighbor search} instead of top-k search is employed. The fixed-radius near neighbor search is amenable to the threshold-based match offered by the TCAM implementation, and reduces the total number of required operations. Algorithm-level evaluation will be presented in Sec.~\ref{sec:evaluation} to study the effects of the adjustment on accuracy. We use a 256 LSH signature length which requires 2 CMAs to store a single entry. 

\subsection{RecSys Operation Mapping} \label{method:mapping}


Given the \fw architecture, careful mapping of the computation, how in Fig. \ref{fig:Algorithm} to \fw is required in order to design the control sequence properly. The crossbars stored the trained weights of the DNN stack and the ItETs and UIETs store the trained \ets for the sparse feature vectors. \chesca{Fig.~\ref{fig:archmapping} illustrates such a mapping where the labels (1a)-(2d*) indicates. We first discuss the operations in the filtering stage.} 
\chesca{\textbf{(1a)} The sparse features are sent to the corresponding \ets, i.e., UIETs and ItET for lookups and pooling. The embedding vectors of the features are obtained by looking up the stored \ets in the ItET CMAs and UIET CMAs using the RAM mode of the CMA. The retrieved embeddings are then aggregated (indicated by \textbf{(1b*)} in Fig. \ref{fig:archmapping}) by the in-memory adder, intra-mat and intra-bank adder trees. \textbf{(1b)} the dense features are sent to the pre-trained filtering sparse feature DNN stack which is implemented with the crossbar arrays. \textbf{(1c)} All features are then sent to the filtering DNN. The output of the filtering DNN is a user embedding vector ($u_{i}$ in Fig.~\ref{fig:archmapping}). \textbf{(1d)} The user embedding vector is then sent to the ItET to retrieve the $N$ nearest neighbors as candidate items inside the ItET CMAs.  The indices of these retrieved (i.e., candidate) item embeddings are then stored in the item buffer (See \textbf{(1d*)} inf Fig. \ref{fig:archmapping}). }

\chesca{We now discuss the operations in the ranking stage. \textbf{(2a)} Each candidate item in the item buffer is then analyzed by the next steps with respect to the user's preferences. \textbf{(2b)} By using the item indices in the item buffer, their corresponding item embeddings are retrieved from the stored \et in the ItET and the ranking embeddings are retrieved in the stored \et in the ranking UIETs using the RAM mode. Note that some UIETs used in the filtering mode can be shared with the ranking stage. The item embeddings are pooled with the ranking embeddings either by concatenation or by an ADD operation using the in-memory adder, intra-bank and inter-bank adder trees. (See (2b*) in Fig. \ref{fig:archmapping}) The output forms a new set of embedding features. Together with the dense features are fed to the ranking DNN stack implemented in crossbars. \textbf{(2c)} the dense features are again obtained from the trained ranking DNN stack. \textbf{(2d)} The remaining crossbar arrays implement the ranking DNN with the pooled feature embeddings as input and return the click-through-rate (CTR) to the CTR buffer. The CTR buffer is a CMA that stores the CTR for each candidate item and the item index which are used for selecting the final top-$k$ items. \textbf{(2e)} The CTR buffer then performs a top-k operation using the threshold match mode of the CMA by searching a vector of all 1's (the maximum allowable CMA input).}

\section{Evaluation}
\label{sec:evaluation}


We evaluated the \rs implementations with two \rs instances: (1) YoutubeDNN model~\cite{covington2016deep} on the MovieLens 1M dataset, 
which includes both the filtering and ranking stage; (2) DLRM model~\cite{DLRM19} targeting at the ranking stage on the Criteo Kaggle dataset. The configurations of the two \rs shown in Table~\ref{tab:conf} were implemented on Nvidia RTX 1080 GPU. We used the tools Nvidia-smi and lineprofiler to obtain energy and latency, respectively. We only compared with the GPU evaluation as other accelerators use older \rs models and datasets.

\dayane{We dimension $B$, $M$ and $C$ based on the largest dataset used in our evaluation (i.e., the Criteo Kaggle) with the configuration shown in Table.~\ref{tab:conf}.} \chesca{Each \et have different number of entries. In this configuration, the maximum size of the \ets in the Criteo Kaggle is 30,000 entries. Since each CMA has 256 rows, 118 CMAs are required to store the embedding table.} \dayane{The number of arrays is rounded up to the nearest power-of-two value, i.e., 128. We choose $C$=32, which corresponds to 4 mats ($M$=4) working in parallel inside each bank, to have a balance between storing small and large \ets. The outputs from the 4 mats are accumulated at the bank level.} \notes{Finally, the Criteo Kaggle dataset has 26 sparse features for ranking. Hence, we  dimension \fw with 32 banks ($B$=32) to accomodate all these features.} The other features are also mapped accordingly, with some mats and CMAs deactivated in a bank according to the size of the \et. \chesca{We use 26 activated banks, 104 activated mats and 2860 activated CMAs for the Criteo Kaggle dataset}.

\dayane{For the MovieLens dataset (also used in our evaluation), due to the much smaller number of rows per ET, we are still able to use the same architecture while keeping idle arrays deactivated.} \chesca{The MovieLens dataset uses 5 UIET for the filtering stage and 6 UIET for the ranking stage (Table \ref{tab:conf}), 5 of which are shared between the filtering and ranking stages. \ets have a maximum of 6040 entries and a minimum of 3 entries. We use 7 active banks, 8 active mat and 54 active CMA in the MovieLens dataset.}

\begin{table}[tb]
\centering
    \caption{\rs configurations and memory mapping on \fw}
    \vspace{-0.25cm}
    \label{tab:conf}
\scalebox{0.95}{\begin{tabular}{|c|c|c|c|c|}
\hline
 & \multicolumn{2}{c}{\textbf{Movielens}} & \multicolumn{2}{|c|}{\textbf{Criteo Kaggle}} \\ \hline
Model & Youtube & Youtube & \multicolumn{2}{c|}{DLRM} \\ \hline
Stage & Filtering & Ranking & \multicolumn{2}{c|}{Ranking} \\ \hline
DNN Network & 128-64-32 & 128-1  & 
\begin{tabular}[c]{@{}c@{}}Bottom MLP \\ 256-128-32 \end{tabular}  & \begin{tabular}[c]{@{}c@{}}Top MLP \\ 256-64-1\end{tabular} \\
\hline
\# UIET (Shared) & 5 (5) & 6 (5) &  \multicolumn{2}{c|}{26} \\ \hline
\# ItET & \multicolumn{2}{c|}{1} &  \multicolumn{2}{c|}{/} \\
\hline
\# Row per ET & \multicolumn{2}{c|}{3000}  & \multicolumn{2}{c|}{28000} \\

\hline\hline
\# Bank &  \multicolumn{2}{c|}{7} &  \multicolumn{2}{c|}{26} \\ \hline
\# Mat &  \multicolumn{2}{c|}{8} &  \multicolumn{2}{c|}{104}  \\ \hline
\# CMA &  \multicolumn{2}{c|}{54}  &  \multicolumn{2}{c|}{2860} \\ \hline
\end{tabular}}
 \vspace{-0.2cm}

\end{table}

\begin{table}[tb]
    \centering
    \caption{Array-level evaluation of CMA, adder-trees and crossbars. \notesdone{Add crossbar values}}
    \vspace{-0.25cm}
    \label{tab:array}
\scalebox{0.92}{\begin{tabular}{|c|c|c|c|}
\hline
\textbf{Component}  & \textbf{Operation} & \textbf{Energy (pJ)} & \textbf{Latency (ns)} \\ \hline

               & Write  & 49.1 & 10.0 \\ 
256$\times$256 & Read & 3.2   & 0.3   \\ 
CMA            & Addition & 108.0  & 8.1 \\ 
               & Search & 13.8  & 0.2 \\ 
\hline

Intra-mat adder tree & 256-bit Add & 137.0  & 14.7 \\ \hline

Intra-bank adder tree & 256-bit Add & 956.0  & 44.2 \\ \hline

\begin{tabular}[c]{@{}c@{}}256$\times$128 Crossbar \end{tabular} & MatMul & 13.8 & 225.0 \\ \hline

\end{tabular}}
\vspace{-2ex}
\end{table}
\subsection{Array-level Evaluation}

We have designed the complete circuit of a 256$\times$256 FeFET-based CMA, and simulated it in HSPICE by employing a Preisach based model for FeFETs \cite{ni18} along with the CMOS Predictive Technology Model (PTM) from \cite{cao2002predictive} with a 45nm technology node. \dayane{In our evaluation, besides the memory cells inside each CMA, we also consider all the peripherals depicted in Fig. \ref{fig:archmapping}(c). The adder trees and communication network are implemented in Verilog and synthesized with Cadence Encounter RTL Compiler v14.10, with the NanGate 45nm open-cell library \cite{knudsen2008nangate}.} The crossbars are evaluated by the Neurosim tool \cite{chen2018neurosim} using a 45nm FeFET model. Table~\ref{tab:array} summarizes the array-level figures-of-merit (FoM) for the different types of accesses supported by the CMA. Table~\ref{tab:array} also includes the FoM for a 256$\times$128 crossbar. These values are used for higher-level evaluations.

\subsection{Accuracy Evaluation} \notesdone{FL: You keep saying accuracy here but you used hit rate}
We examined the algorithm-level performance (i.e., accuracy) of \rs when using quantized data representation and when using different distance functions. We implemented a YoutubeDNN filtering model~\cite{covington2016deep} on the MovieLens 1M dataset~\cite{harper2015movielens}, where a FAISS-based distance search is used. We use the hit rate (HR), the \# of hits (i.e., correct predictions) divided by the \# of test users, as the accuracy metric. Three configurations are tested: (1) 32-bit floating-point (FP32) representation and cosine distance (2)8-bit Int and cosine distance (3) 8-bit Int and LSH-based Hamming distance and achieve HR to be 26.8\% / 26.2\% /20.8\%, respectively. \fw incurs around 5.4\% 
accuracy degradation, which indicates that the distance function plays an important role in the accuracy. However, since the filtering stage only provides a coarse selection of the candidate items, such accuracy loss is tolerable as the accuracy is retained by the ranking stage.



\subsection{Energy and Latency Evaluation}

We estimated the energy and latency of \fw based on the mapping and the simulated array-level FoM. As the latency and energy improvement of TCAM arrays and crossbars are well studied in previous work~\cite{ni2019ferroelectric}~\cite{chen2018neurosim}, in this section we mainly focus on our findings on ET lookup operations, which is also a bottleneck of the \rs. We compare the latency and energy of TCAM-based LSH search with the LSH search on GPU as well as the original cosine search on GPU. Finally we report the end-to-end system comparison.

\begin{table}[]
    \centering
    \caption{\et operation comparison between the GPU and \fw} 
    \vspace{-0.25cm}
    \begin{tabular}{|c|c|c|c|c|}
    \hline
        \multicolumn{2}{|c|}{Dataset} & \multicolumn{2}{c|}{\textbf{MovieLens}}  & \textbf{Kaggle}\\
    \hline
        \multicolumn{2}{|c|}{Stage} & Filtering & Ranking & Ranking \\
    \hline
        \multirow{3}{*}{\begin{tabular}[c]{@{}c@{}}Latency \end{tabular}} & GPU & 9.27$ \mu$s & 9.60 $\mu$s & 14.97 $\mu$s \\
        
        & iMARS & 0.21 $\mu$s & 0.21 $\mu$s & 0.24 $\mu$s\\
        & \textbf{Speedup} & 43.61$\times$ & 45.17$\times$& 61.83$\times$\\
    \hline
    \multirow{3}{*}{\begin{tabular}[c]{@{}c@{}}Energy \end{tabular}} & GPU & 203.97 $\mu$J& 211.26 $\mu$J & 329.34 $\mu$J\\
        & iMARS & 0.40 $\mu$J & 0.46 $\mu$J & 6.88 $\mu$J\\
        & \textbf{Reduction} & 516.05$\times$ & 458.12$\times$ & 47.90$\times$\\
    \hline

    \end{tabular}
    \vspace{-2ex}
    \label{tab:eval}
\end{table}




\subsubsection{ET lookup operation}

Table~\ref{tab:eval} shows the latency and energy consumption of the ET lookup operation in the two RecSys instances as well as on GPU. All the data are obtained for one item input. 
The ET lookup operation in \fw includes the multiple lookups of CMAs, the intra-mat addition and intra-bank addition. To estimate the latency and energy on \fw, we consider the worst case that all lookups for one ET happen in the same array. Multiple lookups in one array requires multiple read, write and in-memory add operations, incurring higher latency and energy. We also include the latency/energy overhead of communication due to wiring and serialization with the RSC bus/IBC network. Thus, under the aforesaid worst case, the \fw takes 0.24 $\mu s$ latency and 6.88 $\mu J$ energy for a single input on the Criteo Kaggle dataset and achieves 61.83$\times$ latency and 47.9$\times$ energy improvement. For the MovieLens dataset, the \fw achieves 43.1$\times$/45.6$\times$ speedup and 516.05$\times$/458.12$\times$ energy reduction over the GPU counterpart on the filtering/ranking stages. 

From Table~\ref{tab:eval}, it can be seen that on both GPU and \fw, the ranking stage takes more time and energy than the filtering stage for MovieLens because the ranking stage deploys one more ET than the filtering stage as the memory mapping shown in Table~\ref{tab:conf}. Also, the latency and energy for the MovieLens dataset is smaller than another dataset because of the relatively small ET size. These improvements are attributed to the fact that \fw reduces the data movement between the processor and memory by using in-memory ET lookup. Also the adoption of the FeFET technology contributes to part of the improvement.


\subsubsection{NNS operation}
NN Search operations are needed in the filtering stage, and are realized by configuring CMAs to the CAM mode in \fw. The utilization of the CAM search mode enables the NNS operation to be implemented in $O(1)$ time instead of $O(n)$. For the filtering stage on the MovieLens dataset with $O(10^3)$ items, the search latency using the original cosine distance on the GPU is around 13.6 $\mu s$ and it consumes 0.34 $mJ$ for one input. With LSH search with 256 signature length, the GPU spends 6.97 $\mu s$ and 0.15 $mJ$. 
The latency and energy improvement over the GPU counterpart (LSH search) is 3.8e4$\times$ and 2.8e4$\times$ 
as shown in Table~\ref{tab:array}.

\subsubsection{End-to-End}

We compare the end-to-end improvements of \fw over GPU in this section. For the GPU data, we only count DNN stack, ET lookup and NNS operation in the algorithm. For \fw, the ET lookup operation and NNS operation are evaluated as we discussed before. The DNN stack is evaluated using Neurosim~\cite{chen2018neurosim} (FoM shown in Table~\ref{tab:array}), which brings around 2.69$\times$ latency improvement compared to the GPU counterpart. 

For the ranking model on the Criteo Kaggle dataset, \fw achieves 13.2$\times$ latency improvement and 57.8$\times$ energy improvement over GPU.
For the filtering and ranking stage together, \fw achieves 16.8$\times$ and 713$\times$ latency/energy end-to-end improvement on the MovieLens dataset. That is, it can achieve 22025 queries/second compared with the 1311 queries/second on the GPU. The end-to-end improvement is dominated by the ranking stage because each user only goes through the filtering stage once in \fw. However, for each user, the CTR needs to be calculated for each candidate item during the ranking stage.

\section{Conclusion}
\label{sec:conclusion}


We present \fw, an IMC-based accelerator for recommendation systems. \fw uses hierarchical IMC fabric consisting of both crossbars and CMAs to accelerate both the filtering and ranking stages of \rs. We introduce an IMC-friendly embedding table organization and judicious computation mapping to maximize the benefit offered by IMC. \fw achieves 22025 queries/second over 1311 queries/second on the GPU ($16.8\times$ speedup for the MovieLens dataset). Also, $713\times$ end-to-end energy reduction compared with GPU is achieved by \fw.



\bibliographystyle{./my_abbrv.bst}
\bibliography{references}


\end{document}